\documentclass[12pt]{iopart}
\usepackage{iopams}
\usepackage{graphicx}

\begin{document}

\newcommand{\be}{\begin{equation}}
\newcommand{\ee}{\end{equation}}

\title{Johnson noise and  the thermal Casimir effect}

\author{ Giuseppe Bimonte}
\address{%
Dipartimento di Scienze Fisiche Universit\`{a} di Napoli Federico
II Complesso Universitario MSA Via Cintia
I-80126 Napoli Italy;\\ INFN, Sezione di Napoli, Napoli, ITALY\\
} \ead{bimonte@na.infn.it}

\begin{abstract}

We  study the thermal interaction between two nearby thin metallic
wires, at finite temperature. It is shown that the Johnson
currents in the wires give rise, via inductive coupling, to a
repulsive force between them. This thermal interaction exhibits
all the puzzling features found recently in the thermal Casimir
effect for lossy metallic plates, suggesting that the physical
origin of the difficulties encountered in the Casimir problem
resides in the inductive coupling between the Johnson currents
inside the plates. We show that in our simple model all puzzles
are resolved if account is taken of capacitive effects associated
with the end points of the wires. Our findings suggest that
capacitive finite-size effects may play an important role in the
resolution of the analogous problems met in the thermal Casimir
effect.

\end{abstract}

\maketitle

\section{Introduction}

In recent years, great advances in experimental techniques have
stimulated an intense theoretical and experimental activity on the
Casimir effect. By now, the Casimir force has been measured with
an accuracy of a few percent (For  recent reviews on both
theoretical and experimental aspects of the Casimir effect, see
Refs.\cite{bordag}). Apart from the  fundamental interest raised
by this phenomenon, which reveals subtle properties of the quantum
vacuum, Casimir forces  are now finding new and exciting
applications to nanotechnologies \cite{capasso}, to Bose-Einstein
condensates \cite{antezza}, to non-contact atomic friction
\cite{stipe} and to superconductors \cite{bimonte}.

The increasing number of applications of the Casimir effect,
together with the improving level of experimental accuracy reached
by  recent experiments, calls for a corresponding  improved level
of accuracy of theoretical predictions. In his seminal paper,
Casimir considered the case of two infinite plane-parallel plates
made of ideal metal, at zero temperature. In modern experiments,
it is necessary to take account of a number of corrections to this
ideal situation, arising from geometrical factors, like surface
roughness and shape of the plates, finite conductivity of the
plates, and a non-zero temperature. While separate consideration
of all these corrections presented no particular difficulty,
surprisingly it was realized recently that, for the case of
metallic plates, the computation of the combined effect of  finite
metal conductivity and a non-zero temperature poses severe
problems. The purpose of the present  paper is to clarify the
physical origin of these difficulties, and to indicate a possible
resolution. To do this, we consider a simple model of two metallic
wires at finite temperature. We shall see that the inductive
interaction between the Johnson currents in the wires presents all
the puzzling features found in the Casimir problem. For our model,
a simple resolution of these problems can be obtained by taking
account of capacitive effects associated with the end points of
the wires. This suggests that also in the Casimir case the
analogous difficulties can be resolved in a similar way, by
considering capacitive effects resulting from the finite size of
the plates.

To understand the problem of the thermal Casimir effect in real
metals, it is useful to review   briefly the sequence of
successive refinements that have been introduced over the years
into the theory of the Casimir effect for metallic cavities, in
order to describe the effects of temperature and finite
conductivity of the plates, paying attention to the {\it new
physical ingredients} that each step adds to the problem. In the
zeroth order approximation  the plates are considered as ideal
reflectors at zero temperature. Here, the Casimir energy can be
interpreted as arising from the zero-point energy of the
electromagnetic (e.m.) modes supported by the cavity, a purely
quantum phenomenon indeed, that constitutes the genuine essence of
the Casimir effect. Including the effect of temperature  just adds
to the zero-point Casimir energy the contribution of thermally
excited e.m. waves.   In the next step, one includes the effect of
the finite  skin depth of e.m. fields in real metals. For
experimentally relevant distances between the plates, this can be
conveniently done by modelling the metal that constitutes the
plates by the well known plasma model, which assumes a
permittivity $\epsilon_P(\omega)$ \be
\epsilon_P(\omega)=1-{\Omega_p^2}/{\omega^2}\;,\ee where
$\Omega_p$ is the plasma frequency.  Note that
$\epsilon_P(\omega)$ is real, which means that no dissipation is
present (at non zero frequencies). Therefore, the cavity continues
to possess e.m. modes of well defined (real) frequencies, and the
physics remains the same as in the ideal case, apart from the fact
that now the spectrum of eigenfrequencies is altered, and that the
e.m. fields penetrate more or less into the plates. The most
important effect of this refinement is seen at short plate
separations, where it produces a significant decrease in the
Casimir pressure. Inclusion of temperature corrections is
straightforward, of course. Since  the skin depth approaches zero
for infinite plasma frequencies, one expects to smoothly recover
the ideal result, both at zero and non-zero temperature, in that
limit. Detailed computations fully confirm this expectation
\cite{bordag}.   Then comes the next refinement, the inclusion of
dissipation into the picture. Physically, this is an important
issue, because all real metals present ohmic losses.  As far as
the eigenmodes of the cavity are concerned, the presence of a
small amount of dissipation   just entails a slight broadening of
the spectrum (the eigenfrequencies get a small imaginary
component), and therefore one   expects  a small correction to the
Casimir energy (i.e. the zero point contribution). The problem can
be studied quantitatively using the famous Lifshitz theory of
dispersive forces in real media \cite{lifs}. All that is needed is
a model for the metal that includes dissipation, and the simplest
one is of course the classical Drude model $\epsilon_D(\omega)$:
\be\epsilon_D(\omega)=1-\frac{\Omega_p^2}{\omega \,(\omega + i
\gamma)}\;,\label{drude}\ee where $\gamma$ is the relaxation
frequency, that accounts for ohmic losses. If this model is
plugged into Lifshitz theory nothing much happens, at $T=0$, and
one finds, as expected,  small deviations from the plasma model
case.  An unexpected result was obtained when the analysis was
extended to finite temperature \cite{sernelius}, for then large
deviations from the plasma model were found, to the extent that
the Casimir pressure is almost halved for large separations. A
puzzling fact is that the large effect persists also for
vanishingly small amounts of dissipation, i.e. for infinitesimal
$\gamma$'s. The existence of this   discontinuity at $\gamma=0$
attracted a lot of interest by the Casimir community, and
stimulated a large number of studies aiming at either disproving
it or supporting it.

An important progress was made in Refs. \cite{lamor}, where the
problem was investigated using an alternative mathematical
formulation of Lifshitz theory, that allows a clean separation of
the contribution of thermal photons from that of zero-point
fluctuations. In this way, it was realized that the presence of
dissipation has little influence on zero-point fluctuations, and
that the large deviations from the plasma model discovered in
\cite{sernelius} originated from thermally excited evanescent TE
photons that   give rise to an additional {\it   repulsive force}
between the plates. This thermal contribution is present only for
$\gamma \neq 0$, it persists in the limit $\gamma \rightarrow 0$
and at large separations and/or temperatures, it compensates to a
large extent for the zero-point Casimir pressure.

At present, the controversy still goes on, with no clear cut
answer. For a discussion of alternative points of view, the reader
may consults two recent papers by leading experts of the field
\cite{most2, milton}. We just remark that a serious objection
against the result of \cite{sernelius} was raised in
Ref.\cite{romero}, where it was shown that the approach of
Ref.\cite{sernelius} implies a violation of Nernst heat theorem,
if the plates are regarded as perfect lattices. However, this
result too has been subjected to criticism (See \cite{milton} and
Refs. therein). In particular, it has been remarked that
thermodynamic inconsistencies might be overcome if space
dispersion is considered \cite{sveto, sernelius2}. That
space-dispersion should be taken into account is clear, because
the anomalous skin effect becomes important at low temperature.
However, it appears that consideration of space-dispersion is not
sufficient to provide a complete resolution of thermodynamic
problems, because while it ensures that the entropy vanishes in
the limit of zero temperature, it does not avoid the problem of
negative entropies at intermediate temperatures \cite{sveto}.

Survey of the literature gives the feeling that the real {\it
physical} reason of the   puzzling thermal effects brought about
by the presence of dissipation in conductors has not yet been
understood.  The previous discussion suggests than {\it when we
deal with metals, or more in general with conductors, the
simultaneous presence of dissipation and of a finite temperature
brings into the problem a qualitatively new phenomenon, that is
absent either when dissipation is strictly zero, or when the
temperature is zero}. If such a phenomenon  exists, the question
arises whether it is adequately accounted for in existing
treatments of thermal Casimir effect. Indeed the phenomenon in
question is known since a long time: Johnson noise \cite{john}. As
it is well known, Johnson noise is the electronic noise generated
by thermal agitation of charge carriers inside a conductor at
equilibrium. This really is a new physical ingredient of the
problem, which has nothing to do with the physics of cavity modes
that exist in a lossless cavity. It is well known that near-field
effects associated with Johnson noise are important, for example,
for the problem of stability ion-traps \cite{vuletic}. The
relevance of Johnson  noise for the thermal Casimir effect  is
discussed in the next Section, with the help of a simple system of
two conducting wires.

\section{Thermal interaction between two wires}

As we pointed out in the Introduction, the enigmatic features of
the thermal Casimir effect arise from the existence of a thermal
{\it repulsive} force, that is present between a pair of metallic
plates at finite temperature,  when the latter are described by
the Drude model. A detailed analysis \cite{lamor,esquivel} of the
real-frequency spectrum of the thermal interaction between two
parallel plates at separation $a$, as given by Lifshitz theory,
revealed that the repulsive force is due to  TE evanescent thermal
fluctuations of the  e.m. field inside the cavity, with  typical
(real) frequencies of order $\tilde{\omega}=\gamma
(\omega_c\,/\Omega_p)^2$, where $\omega_c=c/a$ is the
characteristic frequency of the cavity. For typical Casimir
experiments, and even more so at low temperature, $\tilde{\omega}$
is much less than $\omega_c$. At such low-frequencies, retardation
effects are negligible and the relevant e.m. fluctuations
basically consist of quasi-static magnetic fields. The physical
origin of the resulting interaction between the plates can be
clearly understood using Rytov's theory \cite{rytov} of e.m.
fluctuations, which is at the basis of Lifshitz theory of
dispersion forces. According to Rytov's theory, e.m. fluctuations
outside material bodies are produced by microscopic fluctuating
currents, which are present in the interior of any absorbing
medium, according to the fluctuation-dissipation theorem.  When
the problem is  analyzed from this point of view, one realizes
that the repulsive force, mediated by quasi-static magnetic
fluctuations, arises from the inductive coupling between the
Johnson currents, that exist inside the plates. To gain further
confidence in the correctness of this simple physical picture, it
is a good strategy to devise a simple model which only involves
this ingredient of the problem, and see what happens. For this
purpose, we consider a system of two identical pieces ${\cal C}_1$
and ${\cal C}_2$ of thin metallic wire at temperature $T$,
displaced by an amount $\vec{a}$ from each other (we consider the
orientations of the wires as fixed once and for all, and therefore
their mutual position   is determined by the displacement
$\vec{a}$  connecting an arbitrary point $O_1$ of ${\cal C}_1$ to
an arbitrary point $O_2$ of ${\cal C}_2$), and we ask what force
they exert on each other,  as an effect of the respective Johnson
currents. Determining this force for two thin wires is indeed much
simpler than for two bulk plates. As we are dealing with a
quasi-static problem, we can neglect retardation effects and we
can regard the force as arising from a direct instantaneous
interaction between the currents $i_1$ and $i_2$ in the wires.
Moreover, we can safely assume that the currents $i_1$ and $i_2$
depend only on time, and not on the position  along the wires
(charge fluctuations decay within a typical time
$\tau=\gamma/\Omega_p^2$, which is very short compared with the
relevant low frequencies of the problem). Under these simplifying
conditions, a simple method to study the interaction between the
wires, completely equivalent to Rytov-Lifshitz theory, consists in
replacing the noisy wires by "ideal" noiseless wires, whose
end-points are connected to noise e.m.f. generators. Then, our
system of two wires is described by the following equations:
\begin{eqnarray}
{\cal L} \,\frac{d i_1}{dt}+{\cal M}(\vec{a})\frac{d i_2}{dt}+R\,
i_1 &=& {\cal E}_1(t)\;, \nonumber\\
{\cal L}\, \frac{d i_2}{dt}+{\cal M}(\vec{a})\frac{d i_1}{dt}+R\,
i_2 &=& {\cal E}_2(t)\;.\label{sys}
\end{eqnarray}
In these equations ${\cal L}$ and $R$,   are, respectively, the
self-inductances and the resistances of the two wires, ${\cal
M}(\vec{a})$ is the mutual inductance, and ${\cal E}_i(t)$  is the
random e.m.f. in the wire ${\cal C}_i$. In principle, ${\cal L}$,
${\cal M}$ and $R$ are frequency-dependent quantities, because
both the conductivity of the wires and the skin-depth depend on
the frequency. However, to keep things simple, we shall neglect
this complication and work with constant inductances and
resistances, which is correct for sufficiently thin wires and at
low enough frequencies. Note that the only quantity that depends
on the separation $\vec {a}$ is the mutual inductance ${\cal
M}(\vec{a})$. As it is well known \cite{feyn2}, in general \be
{\cal M}^2 \le {\cal L}^2 \;. \ee The power spectrum \footnote{The
Fourier transform $g(\omega)$ of a function $f(t)$ is normalized
here such as $g(\omega)=\int_{-\infty}^{\infty} dt \, f(t)$.} of
the random e.m.f.'s is   \cite{nyq} \be \langle {\cal
E}_i(\omega)\,{\cal E}_j^*(\omega')\rangle=4 \pi k_B
T\,R\,E(\omega/\omega_T)\,
\delta(\omega-\omega')\,\delta_{i\,j}\;,\label{spec}\ee where
angle brackets denote statistical averages, $i,j=1,2$,
$\omega_T=k_B T/\hbar$ and $E(y)=y\,(e^y-1)^{-1}$ \footnote{This
form of the spectrum is valid for wires that are not exceedingly
thin \cite{turl}}. The force ${\vec F}_{12}(\vec{a})$ on the wire
${\cal C}_2$ can be written as \cite{jacks}: \be {\vec
F}_{12}(\vec{a})=\langle i_1 \,i_2 \rangle\, \vec{\nabla}_a {\cal
M}(\vec{a})\;.\ee The correlator $\langle i_1 \,i_2 \rangle$ can
be easily computed by taking the time-Fourier transform of
Eqs.(\ref{sys}) and using Eq. (\ref{spec}). After some
computations, we obtain for the force the simple formula: \be
{\vec F}_{12}= -k_B T\,H\,\vec{\nabla}_a (m^2)\;,\label{force}\ee
where $m= {\cal M}/{\cal L}$ and $H$ is the quantity: \be
H=\frac{1}{\pi} \int_0^{\infty} d \omega\,\omega E\left(\frac{
\omega}{\omega_T}\right)\,{\rm Im}\,\left[(\omega_R-i
\omega)^2+\omega^2 m^2\right]^{-1}\;,\label{H}\ee   where
$\omega_R=R/{\cal L}$. As we see, the frequencies $\omega$ that
contribute to the $H$ are clearly in the range $0<\omega <
\min\{\omega_R,\omega_T\}$, and therefore the low-frequency
approximation made in Eqs.(\ref{sys}) is justified if either
$\omega_R$ or $\omega_T$ are low enough.

Let us consider now the features of the force. It is easy to see
that $H$ is positive definite, and therefore, since ${\cal M}^2$
decreases as $a$ increases, the force is  {\it repulsive}.
Moreover, we note that the force vanishes if we take $R=0$, i.e.
ideal inductances. Both features are analogous to what is found in
the thermal correction to the Casimir pressure \cite{lamor}. Let
us go ahead and check the zero-resistance limit. By making the
change of variables $x=\omega/\omega_R$ in the integral in Eq.
(\ref{H}), it is easy to verify that $H$ is only a function of
$m^2$ and $\omega_R/\omega_T$. However, the dependence on the
latter quantity is only via the Boltzmann factor $E(x
\,\omega_R/\omega_T)$, and since  $x$  is of order one or less, we
see that for small $R$'s, $E(x \,\omega_R/\omega_T)$ becomes one,
which represents the classical value. Therefore, in this limit $H$
is a function only of $m^2$, $H= f(m^2)$, and we obtain \be
\lim_{R \rightarrow 0}{\vec F}_{12}= -k_B
T\,f(m^2)\,\vec{\nabla}_a (m^2) \, \;.\label{lowR}\ee
As we see, the force does not vanish for $R\rightarrow 0$. This is
exactly the same kind of discontinuity that occurs in the Casimir
case with lossy plates, for vanishing dissipation.

\section{Thermodynamic features of the interaction}

To verify if the analogy with the Casimir case is complete,  we
now consider the thermodynamic features of the interaction between
the wires. For this purpose, we need to compute the corresponding
free energy ${\cal F}$. Since ${\vec F}_{12}=-\vec{\nabla}_a {\cal
F}$, we obtain form Eq.(\ref{force}) and Eq.(\ref{H}) \be {\cal
F}=\frac{k_B T}{\pi} \int_0^{\infty}\frac{d \omega}{\omega}\, E
\left( \frac{\omega}{\omega_T}\right)\,{\rm Im}\log
\left[1+\left(\frac{\omega \,m}{\omega_R-i
\,\omega}\right)^2\right]\,.\label{freen} \ee Let us consider now
the low temperature limit. If the wires have no impurities, at
liquid Helium temperatures and below, the resistance $R(T)$
approaches zero as $T^2$ \cite{kittel}. Therefore near $T=0$, we
always have $\omega_R \ll \omega_T$, and reasoning as before, we
can substitute to $E(\omega/\omega_T)$ its classical value, i.e.
$H=1$. But then the integrand in Eq. (\ref{freen}) becomes
independent on $T$ and we find that the free energy is of the form
\be {\cal F} \approx g(m^2)\,k_B T \; \ee where $g(z)$ is a
positive function (because the imaginary part of the argument of
the logarithm in Eq. (\ref{freen}) is  positive definite).
Recalling that the entropy $S$ is $S=-\partial {\cal F}/\partial
T$, we find: \be\lim_{T \rightarrow 0}S=- k_B\, g(m^2) \;\equiv
S_0 <0.\label{entr}\ee Since $S_0$ depends on the separation among
the wires through $m^2$, this result represents a violation of
Nernst heat theorem. Again, this is exactly analogous to what is
found in the Casimir case. Therefore, our simple model   strongly
suggests that the troubles with the thermal Casimir effect
originate from the Johnson noise in the plates.
\begin{figure}
\includegraphics{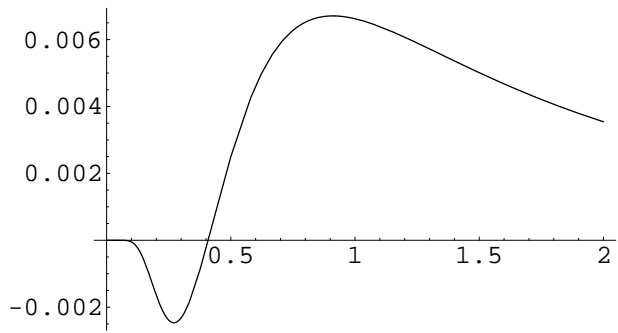}
\caption{\label{plotssigma} Plot of the   free energy (in units of
$\hbar \omega_C$) as a function of $t=k_B T/(\hbar \omega_C)$. See
text for details.}
\end{figure}

The question then is:  have we missed anything here? At a closer
inspection, we see that we omitted an important ingredient. We did
non consider that the wires have a finite length, and therefore
have end points, where charges can accumulate and produce
capacitive effects. A simple way to account for the effect of the
end-points in our elementary approach is to include a capacitance
$C$ in the impedance $Z$ of the wires, such that $Z=R-i\,\omega
\,{\cal L}+i/(\omega\, C)$. We might consider also the possibility
of a mutual capacitance among the wires, but to keep things simple
we shall suppose that the orientation of the wires is such as to
minimize this effect (for example, the wires could be at right
angles with each other, so that the respective end-points are far
from each other). It is clear that the presence of the
capacitances makes a big difference at low frequencies, which are
the source of the troubles, since they work as "high-pass filters"
that block low frequencies. The only modification introduced by
the capacitances in the previous formulae is that the quantity
$(\omega_R -i\, \omega)$ occurring inside the square brackets in
Eqs. (\ref{H}) and (\ref{freen}) gets replaced by $\omega_R -i
\omega + i\, \omega_C^2/\omega$, where $\omega_C=1/\sqrt{{\cal L}
\,C}$. For a straight wire of length $L$, $\omega_C$ is expected
to be of order $2 \pi c/(2 L)$, i.e. the resonance frequency of a
linear antenna.

Consideration of the capacitances resolves indeed all the
problems.  First, we have verified that the force now vanishes in
the zero resistance limit, and thus we recover the ideal metal
result. Moreover, we found that even for finite values of the
resistance, the value of the force may be affected in an important
way, but we shall not discuss this in detail here. What we
consider more at length is instead the low temperature limit of
the free energy. Since $\omega_C$ is independent of $T$, at
sufficiently low temperatures, we always have (for resistors with
no impurities) $\omega_R \ll \omega_T \ll \omega_C$. In this
limit, it can be verified that \be {\cal F}=- \frac{16
\pi^5\,m^2}{63}\,\left(\frac{k_B T}{\hbar \,\omega_C}\right)^6
\,\hbar\, \omega_R\;.\ee Obviously, the entropy is positive, and
vanishes as $T\rightarrow 0$, as required by Nernst theorem.
However, numerical computations show that the entropy is {\it
negative at intermediate} temperatures. This can be seen from the
plot of the free energy (in units of $\hbar\, \omega_C$) as a
function of the reduced temperature $t=k_B T/(\hbar\, \omega_C)$
in Fig.1. The curve was computed by taking $ m=0.8$ and
$\omega_R(t)=5 \,t^2\,\omega_C$. As we see, there is a region of
temperatures $t$ where the slope of the curve is positive, which
corresponds to a negative entropy. This is not necessarily a
problem, though, because what needs to be positive is the {\it
total} entropy of the system, which includes the self-entropies of
the wires. Each wire, being now an $RLC$ oscillator, has a free
energy ${\cal F}_{\rm self}$ equal to: \be {\cal F}_{\rm self}=k_B
T\,\log[1-\exp(-\hbar \omega_C/(k_B T))]\;.\ee As we have checked
numerically, inclusion of the wires self-entropies   makes the
total entropy of the system positive at all temperatures (we could
not obtain an analytical proof of this). Therefore, the inclusion
of capacitive effects related to the edges of the wire resolves
all thermodynamical inconsistencies as well.

\section{Conclusions}

The  results of this paper  suggest that the controversial thermal
correction to the Casimir interaction between  lossy  conductors,
has its physical origin in the inductive coupling among the
Johnson currents existing inside the conductors. This simple
physical picture is suggested by Rytov's  theory of e.m.
fluctuations, which is at the basis of Lifshitz theory of
dispersion forces. To verify its correctness,   we  studied the
force arising between two thin metallic wires, as a result of
Johnson noise. We found that the interaction displays the four
basic features found in the thermal Casimir problem, namely:
\begin{itemize}
    \item  it is repulsive;
    \item it persists in the zero resistance limit;
    \item it is absent in the case of strictly dissipationless
    wires;
    \item it violates the Nernst heat theorem.
\end{itemize}
These results appear to us as a clear indication of the important
role played by Johnson noise in the  thermal Casimir effect for
metallic bodies.

In the simple case of two wires studied in this paper, we have
shown  that all the puzzles raised by this interaction can be
resolved by considering capacitive effects arising from the wires
finite size, which both ensure smooth convergence to the ideal
case in the limit of zero resistance, and resolve as well all
thermodynamic inconsistencies at low temperature.

The relevance of capacitive edge effects for resolving the
analogous difficulties met in the thermal Casimir effect requires
further investigations. Here we just content ourselves with a few
general remarks.  Obviously, capacitive effects   are expected to
be important only if the relevant thermal current-fluctuations
have a typical spatial size that is comparable to  the plates size
$L$.  Recent investigations of the thermal corrections to the
Casimir force \cite{lamor, esquivel} show that, at room
temperature, the troublesome thermal fluctuations are associated
with evanescent TE modes, with characteristic frequency of order
$\tilde{\omega}=\gamma \, c^2/(\Omega_p^2\,a^2)$, and
characteristic spatial size of the order of the plate separation
$a$. Therefore, we expect that capacitive finite-size effects will
be important, for plate separations $a$ not too   smaller than the
plates size $L$. This is the typical case with micro-mechanical
devices, that are currently under intense investigation
\cite{capasso}.

At low temperature, the situation is more complicated.  As the
temperature is lowered in the cryogenic range, the increasing
degree of spatial correlation between the  Johnson currents,
implied by the anomalous skin effect, leads to a gradual
suppression of fluctuations at small scales. It is conceivable
that at very low temperatures, independently of the separation
$a$, the current fluctuations become so  correlated as to have a
spatial extent comparable with the size of the plates. When this
point is reached, edge effects enter into play and may become
essential for a correct description, as discussed in this paper.

\end{document}